\documentclass[preprint]{aastex63}
\usepackage{lineno}

\usepackage{graphicx,epsfig,fancyhdr,psfig,rotating,amsmath,epsf,txfonts,natbib,epstopdf,multirow,booktabs}

\received{***}
\revised{***}
\submitjournal{ApJ}

\shorttitle{Decayless oscillations in solar coronal bright points}
\shortauthors{Gao et al.}

\def \kms {{\rm km\;s$^{-1}$}}
\def \arcsec {$^{''}$}

\def \ck {C_\text{k}}

\graphicspath{{./}}

\begin{document}
\title{Decayless oscillations in solar coronal bright points}

\correspondingauthor{Hui Tian}
\email{huitian@pku.edu.cn}

\author{Yuhang Gao}
\affiliation{School of Earth and Space Sciences, Peking University, Beijing, 100871, China}
\affiliation{Centre for mathematical Plasma Astrophysics (CmPA), KU Leuven, Celestijnenlaan 200B bus 2400, B-3001 Leuven, Belgium}

\author{Hui Tian}
\affiliation{School of Earth and Space Sciences, Peking University, Beijing, 100871, China}
\affiliation{National Astronomical Observatories, Chinese Academy of Sciences, Beijing, 100012, China}

\author{Tom Van Doorsselaere}
\affiliation{Centre for mathematical Plasma Astrophysics (CmPA), KU Leuven, Celestijnenlaan 200B bus 2400, B-3001 Leuven, Belgium}

\author{Yajie Chen}
\affiliation{School of Earth and Space Sciences, Peking University, Beijing, 100871, China}

\begin{abstract}
Decayless kink oscillations of solar coronal loops (or decayless oscillations for short) have attracted great attention since their discovery. 
Coronal bright points (CBPs) are mini-active regions and consist of loops with a small size. 
However, decayless oscillations in CBPs have not been widely reported. In this study, we identified this kind of oscillations in some CBPs using 171\,\AA\, images taken by the Atmospheric Imaging Assembly (AIA) onboard the Solar Dynamics Observatory (SDO). After using the motion magnification algorithm to increase oscillation amplitudes, we made time-distance maps to identify the oscillatory signals. We also estimated the loop lengths and velocity amplitudes. We analysed 23 CBPs, and found 31 oscillation events in 16 of them. The oscillation periods range from 1 to 8 minutes (on average about 5 minutes), and the displacement amplitudes have an average value of 0.07 Mm. The average loop length and velocity amplitude are 23 Mm and 1.57 \kms, respectively. Relationships between different oscillation paraments are also examined. Additionally, we performed a simple forward model to illustrate how these sub-pixel oscillation amplitudes (less than 0.4 Mm) could be detected. Results of the model confirm the reliability of our data processing methods.
Our study shows for the first time that decayless oscillations are common in small-scale loops of CBPs. These oscillations 
allow for seismological diagnostics of the Alfv\'{e}n speed and magnetic field strength in the corona.
\end{abstract}

\section{Introduction}
\label{sec:intro}

How the solar corona is heated to a high temperature of one million Kelvin is one of the most important problems in solar physics. 
Magnetohydrodynamic (MHD) wave heating \citep[e.g.,][]{1947MNRAS.107..211A,2015RSPTA.37340261A} and nano-flare heating \citep{1988ApJ...330..474P,2006SoPh..234...41K} are two main explanations for this problem. 
The MHD wave heating mechanism suggests that MHD waves excited by the photospheric convective motions carry energy to the corona and dissipate, thus heating the surrounding plasma \citep[for a review see][]{2020SSRv..216..140V}. 
At the end of the last century, the launch of the Solar and Heliospheric Observatory
\citep[SOHO;][]{1995SoPh..162....1D} and the Transition Region and Coronal Explorer \citep[TRACE;][]{1999SoPh..187..229H} led to the first discovery of the presence of abundant MHD waves in the corona, including slow waves \citep{1998ApJ...505..957C,1999ApJ...514..441O,1999SoPh..186..207B}
and transverse oscillations of coronal loops \citep{1999Sci...285..862N,1999SoPh..187..261S,1999ApJ...520..880A}.

Coronal loops are the main structures in coronal active regions, filled in with high-density plasma frozen on closed magnetic field lines. These loops are important waveguides of coronal MHD waves, with various wave modes discovered in them, such as transverse kink waves, fast sausage waves, and longitudinal slow waves \citep{2020ARA&A..58..441N,2016ApJ...823L..16T,2020SSRv..216..136L,2021SSRv..217...34W}. Standing modes of transverse kink waves, also known as kink oscillations, are found to have two regimes, which are respectively called decaying and decayless oscillations \citep[see review by][]{2021SSRv..217...73N}.

Decaying oscillations have been widely observed in extreme ultraviolet (EUV) bands since 1999 \citep{1999Sci...285..862N,1999ApJ...520..880A}. 
They are usually caused by impulsive external energy release events such as flares and coronal mass ejections (CMEs), with an initial displacement amplitude of a few megameters and a rapid decay over time \citep[e.g.,][]{2013A&A...552A..57N,2016A&A...585A.137G,2018NatSR...8.4471S}. 
The damping mechanisms may be associated with resonant absorption \citep{2002ESASP.506..745R,2002A&A...394L..39G} and development of Kelvin-Helmholtz instability  \citep[e.g.,][]{1983A&A...117..220H,2008ApJ...687L.115T,2014ApJ...787L..22A,2021ApJ...910...58V}.

Decayless oscillations of coronal loops were first discovered by \citet{2012ApJ...751L..27W} and \citet{2012ApJ...759..144T} through imaging and spectral observations, respectively, which are characterized by no obvious damping in multiple oscillation cycles. 
They are found to be ubiquitous in coronal loops \citep{2015A&A...583A.136A}, and not related to any external eruptive events like flares and CMEs \citep[although some CME or flare-induced non-damping oscillations have also been investigated, e.g., ][]{2012ApJ...751L..27W,2020A&A...638A..32Z,2021A&A...652L...3M}. This fact allows these oscillations to be used to diagnose the magnetic field in the corona \citep[e.g.,][]{2012ApJ...759..144T, 2013A&A...552A..57N}. Moreover, they can continually transfer energy from the lower atmosphere to the upper atmosphere, closely relating them to coronal heating \citep[e.g.,][]{2019ApJ...883...20G,2019ApJ...870...55G,2021ApJ...922...60S,2021ApJ...908..233S}.

Statistical studies on decayless oscillations show that their displacement amplitudes range from 0.05--0.5 Mm, with an average of 0.17 Mm \citep{2015A&A...583A.136A}. The velocity amplitudes are approximately 1--8 \kms \citep{2012ApJ...759..144T,2016A&A...591L...5N}. On the other hand, the decaying oscillations have displacement amplitudes of 1--10 Mm and velocity amplitudes above 10 \kms \citep{2016A&A...585A.137G,2019ApJS..241...31N}, which are both significantly larger than decayless oscillations. The periods of decayless oscillations are 1.5--10 min, with an average period of 251 s. The periods are also found to scale with lengths of the oscillating loops \citep{2015A&A...583A.136A}. In addition, the displacement amplitudes also increase with periods and loop lengths, while the velocity amplitudes show no clear correlations with these two \citep{2016A&A...591L...5N}. The harmonic properties of decayless oscillations are also of interest. They generally appear as the fundamental harmonic 
\citep{2013A&A...560A.107A,2015A&A...583A.136A}, while \cite{2018ApJ...854L...5D} found the existence of the second harmonic. 

Many theoretical and simulation works have concentrated on why the decayless oscillations do not damp rapidly like the decaying ones. One simple model is to consider a harmonic driver at the loop footpoints (possibly related to the photospheric 3--5 min oscillations) to maintain the oscillations \citep[e.g.,][]{2013A&A...552A..57N,2017A&A...604A.130K}, but there are some difficulties in interpreting other observational characteristics. \cite{2016A&A...591L...5N} proposed a self-oscillation model, which suggests that the oscillations can be driven by an external quasi-steady flow. The flow may be associated with the  photospheric supergranulation motions at the footpoints \citep{2020ApJ...897L..35K}, or vortex shedding process \citep{2009A&A...502..661N,2019PhRvL.123c5102S} in the corona \citep[modeled by][]{2021ApJ...908L...7K}. \cite{2020A&A...633L...8A} described another model with a random driver at the footpoints, and obtained results that appear to closely match observations \citep[see also][]{2021MNRAS.501.3017R,2021SoPh..296..124R}. Additionally, there are some 3D MHD simulations modeling the oscillations as KHI vortexes enhanced by the resonant absorption process \citep{2016ApJ...830L..22A}. When combining the model with a harmonic driver at the footpoints, the undamped oscillation amplitudes could be well reproduced \citep{2019ApJ...870...55G}.

Although the decayless oscillations have already been well studied in coronal loops, they have not been widely reported in coronal bright points (CBPs).
CBPs are small-scale bright structures in the quiet corona, and can be observed in the EUV and soft X-ray bands \citep[for a review see][]{2019LRSP...16....2M}. High-resolution EUV observations have shown that CBPs are usually composed of many closed loops connecting magnetic elements of opposite polarities, while these loops are much smaller than coronal loops in size \citep{1979SoPh...63..119S}. \cite{2012ApJ...759..144T} first detected decayless oscillations in two CBPs using spectroscopic observations, and obtained their Doppler velocity amplitudes and periods. However, a detailed study using EUV imaging observations is still lacking. 

In this paper, we focus on the decayless oscillations in CBPs at 171\,\AA\, with the Atmospheric Imaging Assembly \citep[AIA;][]{2012SoPh..275...17L} onboard the Solar Dynamics Observatory \citep[SDO;][]{2012SoPh..275....3P}. 
The data we used for the study are described in Section\,\ref{sec:obs}. We present the analysis methods and statistical results in Section\,\ref{sec:ana}, and discuss the results in Section\,\ref{sec:dis}. Finally, our findings are summarized in Section\,\ref{sec:sum}.

\section{Observational data}
\label{sec:obs}

The observation data used in this study are from SDO/AIA. 
AIA can provide high-resolution images of the solar disk in multiple wavelengths at almost the same time, focusing on the corona and transition regions in the solar atmosphere. It has a spatial resolution of 1.5\arcsec, while the pixel size is 0.6\arcsec. The cadence for the EUV wavelengths is 12\,s.
We chose the AIA 171\,\AA\, channel because it can be used to image the low corona and it is very suitable for the study of CBPs. 

We first selected appropriate CBPs from the 171\,\AA\, data. Although there are some automatic methods for CBP identification \citep[e.g.,][]{2001A&A...374..309B,2003ApJ...589.1062H,2015ApJ...807..175A}, we still chose to select them manually, since appropriate CBPs for our study need to meet strict requirements. They should have clear loop structures, and the loops should be relatively stable, which means that the CBPs are not in the stage of formation or disappearance. We examined the data from January 2017 to March 2018, and selected 23 CBPs which can meet our requirements. Their information can be found in Table \ref{CBPs}.

\section{Analysis and results}
\label{sec:ana}
The typical sizes of CBPs are 4--40 Mm \citep{2019LRSP...16....2M}, much smaller than the lengths of oscillating coronal loops \citep[219 Mm on average, according to][]{2015A&A...583A.136A}. Since it is difficult to resolve the loop structures of CBPs and detect low-amplitude oscillations, we adopted a series of analysis techniques to aid our identification of decayless oscillations in CBPs. 

\begin{figure*}[ht!]
	\centering
	\includegraphics[trim=0.0cm 0.8cm 0.0cm 0.5cm,width=1.0\textwidth]{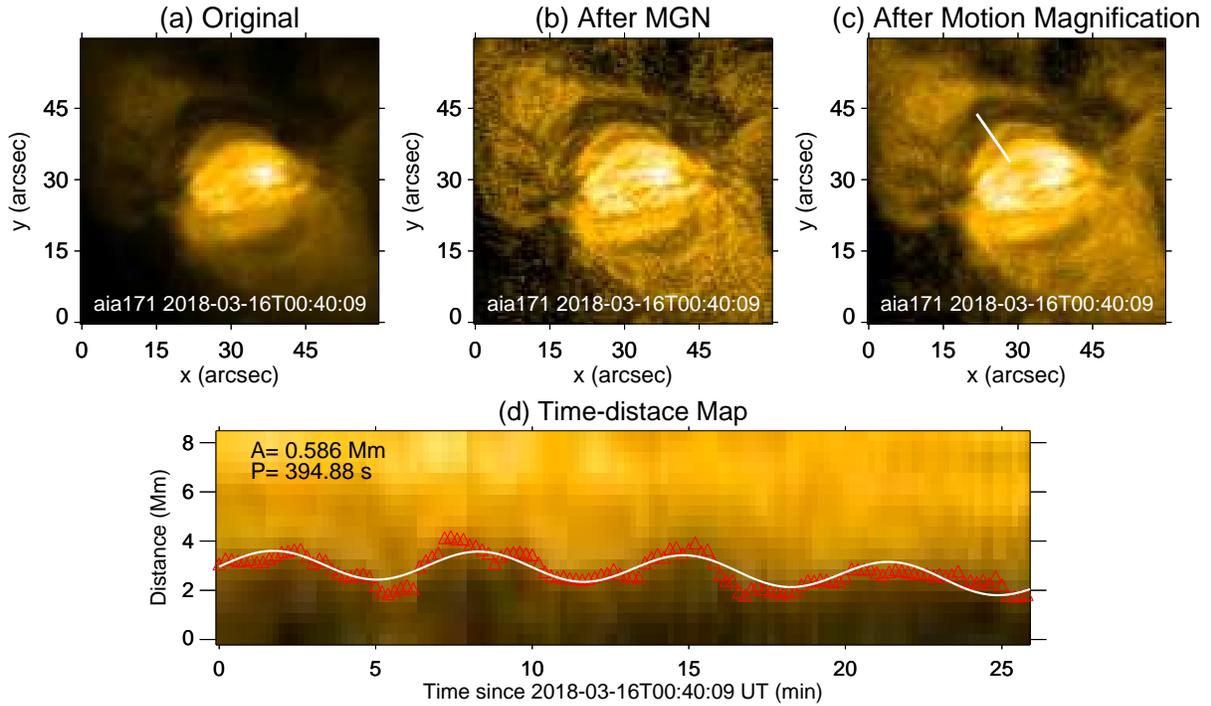}
	\caption{Panel (a) shows a CBP observed in the original AIA 171\,\AA\,\,image. Then the image is enhanced by the MGN method (panel b) and processed by the motion magnification method (panel c). Panel (d) is the time-distance map for the slit marked with a white line in panel (c), with the given distance starting at the left top of the slit.
	The red triangles marked the edge of the oscillating loop. The white curve shows the fitting result. The fitted displacement amplitude ($ A $) and period ($ P $) are indicated in the upper left corner of panel (d).}
	\label{fig:example}
\end{figure*}

We take CBP No.\,22 in Table \ref{CBPs} as an example to demonstrate our methods. Figure\,\ref{fig:example}(a) shows the shape of the CBP in the original AIA image.
Firstly, we used the Multiscale Gaussian Normalization (MGN) method developed by \cite{2014SoPh..289.2945M} to enhance the image and highlight some fine structures. The results are shown in Figure\,\ref{fig:example}(b). The enhanced datacube was then processed by the motion magnification
algorithm \citep{2016SoPh..291.3251A}, which can magnify the transverse displacement amplitudes and retain the original periods. 
This algorithm has been widely used in the study of low-amplitude transverse oscillations of coronal loops \citep[e.g.,][]{2018ApJ...854L...5D,2018A&A...617A..86L,2019ApJ...884L..40A,2021A&A...652L...3M,2021SoPh..296..135Z}. 
In this study, we chose a magnification factor of 5. 
Figure\,\ref{fig:example}(c) presents a frame in the processed datacube, and along the slit marked by the solid white line, we plotted a time-distance map, as shown in Figure\,\ref{fig:example}(d). 
The slit is nearly perpendicular to the loop axis, with a width of 5 pixels. We calculated the average intensity over the width in order to increase the signal-to-noise ratio \citep[see][]{2013A&A...552A..57N,2013A&A...560A.107A,2015A&A...583A.136A}. In the time-distance map, it is clearly seen that there are transverse oscillations lasting for about four cycles  without any apparent damping.

\begin{figure*}[ht!]
	\centering
	\includegraphics[trim=0.0cm 0.8cm 0.0cm 0.5cm,width=0.8\textwidth]{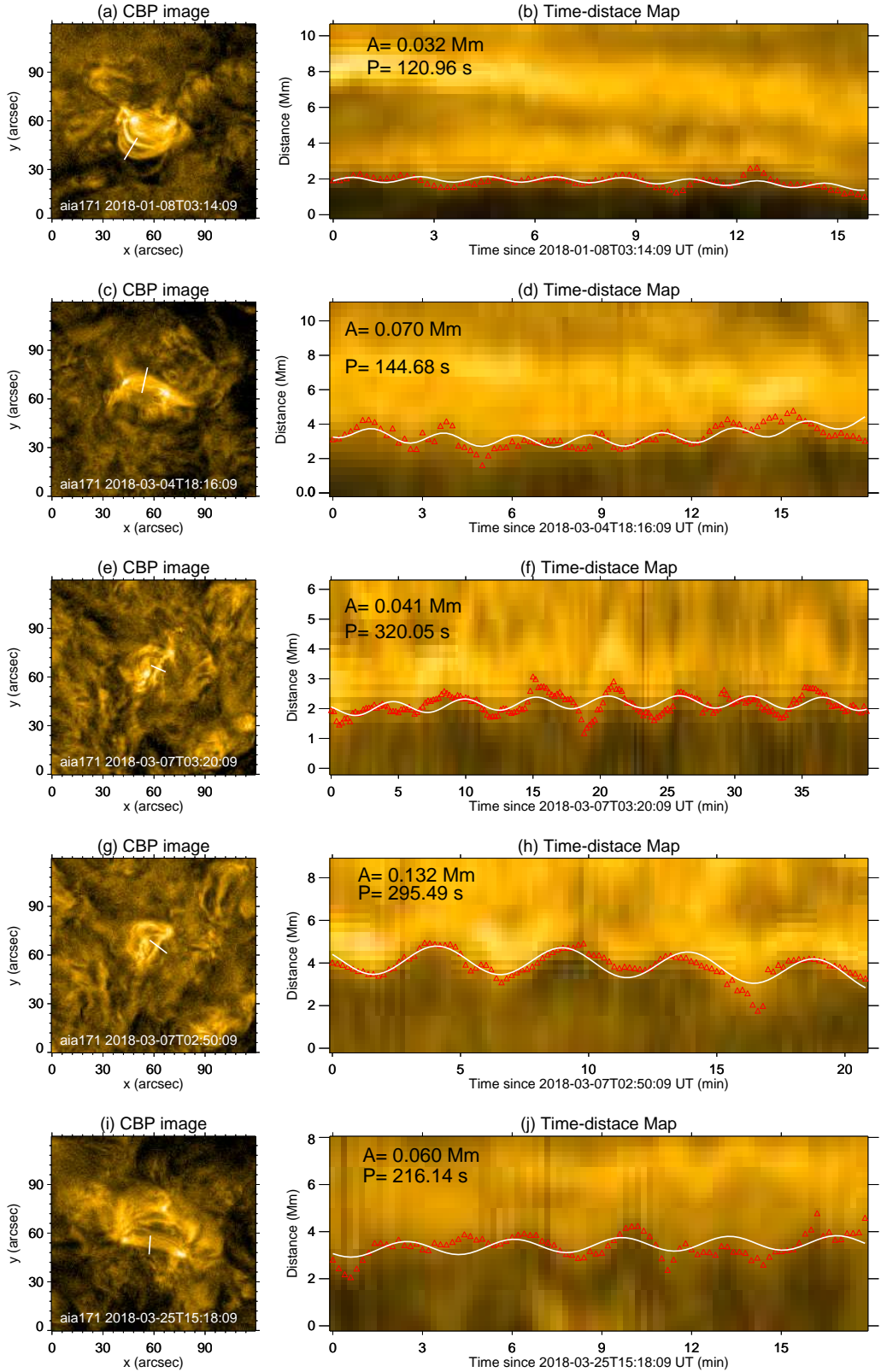}
	\caption{Five decayless oscillation events inside CBPs. The left panel of each row shows the AIA 171\,\AA\, image of the CBP (processed by the MGN method and the motion magnification
		algorithm). The white lines mark the slits for plotting time-distance maps, which are shown in the right panels. Every time-distance map is similar to Figure\,\ref{fig:example}(d), but the displacement amplitudes indicated are 1/5 of the fitting results, which means that they are the actual amplitudes before magnification.}
	\label{fig:events}
\end{figure*}

When studying transverse oscillations of coronal loops, Gaussian fitting is often applied to determine the loop center \citep[e.g.,][]{2012ApJ...751L..27W,2016A&A...585L...6P}.
However, it has been pointed out that there could be many overlapping loop structures in the time-distance map \citep{2013A&A...560A.107A,2015A&A...583A.136A,2016A&A...585A.137G}, which makes it difficult to apply this method. 
In fact, we can see the overlap in Figure\,\ref{fig:example}(d) and Figure\,\ref{fig:events}(d). By assuming that the loops have a constant cross-sectional area, we could use the loop edges to track the oscillations.
The assumption has been supported by a number of observations \citep[e.g.,][]{1992PASJ...44L.181K,2020ApJ...900..167K,2021ApJ...919...47W}.
We determined positions of the loop edge with the methods given in \cite{2013A&A...560A.107A}, i.e., fitting the spatial derivatives of the intensity profile across the loop with a Gaussian function.
The edge positions are marked with red triangles in Figure\,\ref{fig:example}(d).

Finally, we fitted the oscillations with a sine function and a parabolic trend 
\begin{equation}
	a(t)=A\sin(2\pi t/P+\phi)+a_0+a_1t+a_2t^2\,,
\end{equation}
where $ a(t) $ is the displacement at the moment $ t $, $ A $ is the displacement amplitude, $ P $ is the period, and $ \phi $ is the initial phase. Besides, $ a_0, a_1 $, and $ a_2 $ are constant parameters determining the parabolic trend. By using the {\tt\string MPFIT} procedure from the SolarSoft (SSW) package, we could obtain the six paraments $ A, P, \phi, a_0, a_1,$ and $ a_2 $, in which $A$ and $ P $ are the most important ones. In Figure\,\ref{fig:example}(d), the white curve shows the fitting result. The oscillation displacement amplitude and period are 0.586$\pm$0.055 Mm and 395$\pm$5 s, respectively. Considering that the amplitude was magnified by 5 times, the actual amplitude should be 0.117$\pm$0.011 Mm. The velocity amplitude and loop length were estimated as $ V=2\pi A/P $ and $ L=\pi D/2 $ respectively, where $ D $ is the distance between two footpoints of the loop.  For this example, we could obtain a velocity amplitude of 1.88$\pm$0.18 \kms\, and a loop length of 28.9 Mm.
The uncertainty of the velocity amplitude $ \sigma_V $ is obtained with $ \sigma_V^2=\left(\frac{\partial V}{\partial P}\sigma_P\right)^2+\left(\frac{\partial V}{\partial A}\sigma_A\right)^2 $. We did not give the uncertainty of the loop length, since it is hard to determine (see discussions in Section \ref{subsec:cor}).

The same analysis techniques were used for the selected 23 CBPs. As a result, we found that there are 31 oscillation events in 16 of them, and 5 events are shown in Figure\,\ref{fig:events}. All the events have 3--8 oscillation cycles with no apparent damping. We note that there can be more than one oscillation event inside one CBP. Since oscillations are found in 16 of the 23 CBPs, we suggest that decayless oscillations are common in CBPs.

\begin{figure}
	\plotone{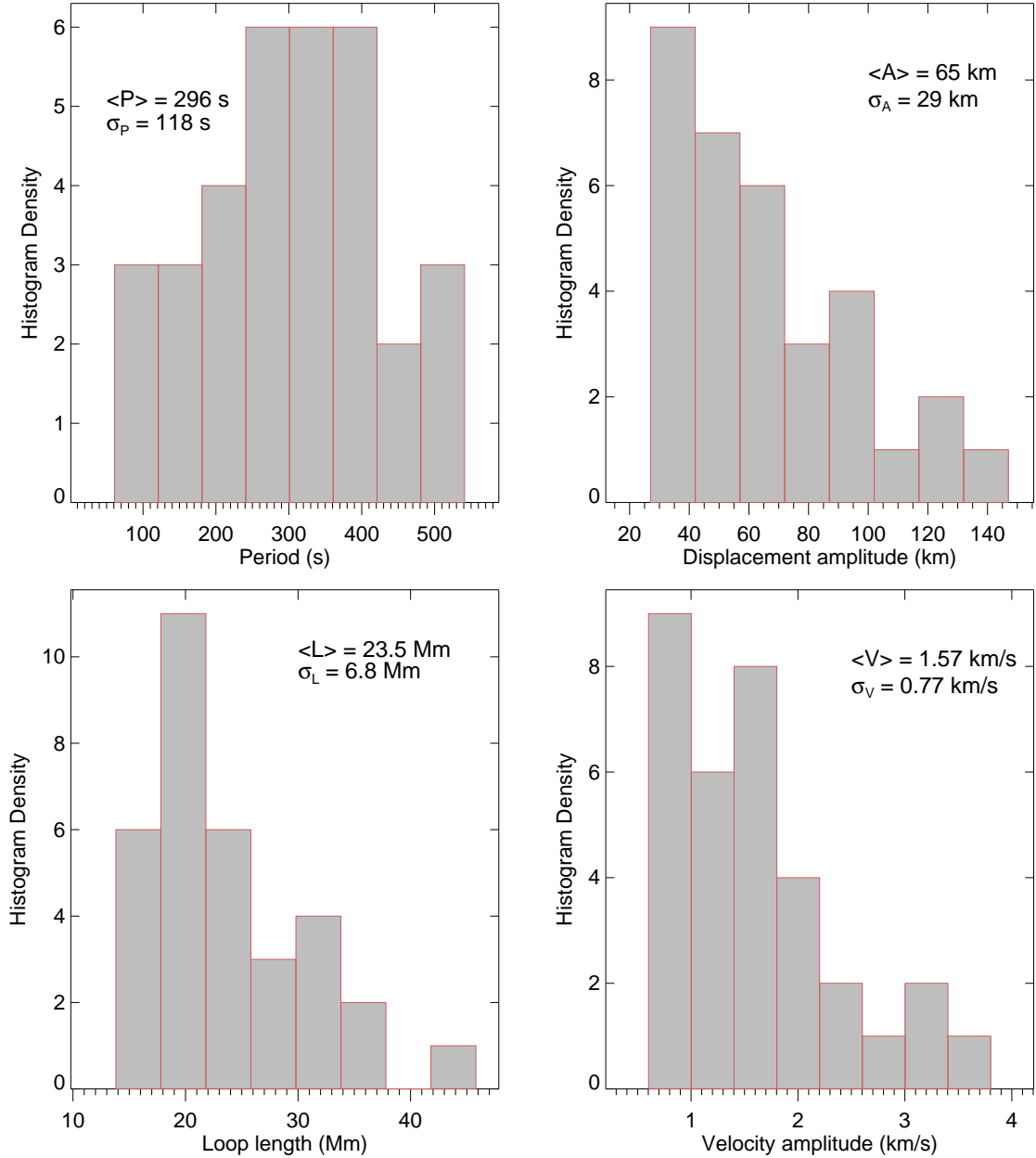}
	\caption{Histograms of oscillation periods, displacement amplitudes, loop lengths, and velocity amplitudes. The average values and standard deviations are indicated in each panel.  \label{fig:histogram}}
\end{figure}

Table \ref{CBPs} lists the four oscillation parameters (oscillation period $ P $, displacement amplitude $ A $, loop length $ L $, and velocity amplitude $ V $) of all the oscillation events. We plotted histograms of these parameters in Figure\,\ref{fig:histogram}. The oscillation periods range from 61 to 498 seconds, with an average of 296\,s. The result is similar to that of oscillating coronal loops, which have a period range of 1.5--10\,min and an average period of 251\,s \citep{2015A&A...583A.136A}.
The estimated lengths of oscillating loops in CBPs are 14--42\,Mm with an average of 23.5\,Mm. For comparison, coronal loops are basically longer than 100\,Mm. 
The displacement amplitudes are 0.027--0.133\,Mm with an average of 0.065\,Mm, which is smaller than those of decayless oscillations of coronal loops (0.17\,Mm on average).
Finally, the velocity amplitudes that we obtained range from 0.6 to 3.6\,\kms. They are smaller than the results from EUV imaging observations of coronal loops \citep{2015A&A...583A.136A,2016A&A...591L...5N}, but close to the spectroscopic observation results (1-2\,\kms) from \cite{2012ApJ...759..144T}. Additionally, we point out that the 2 CBPs investigated by \cite{2012ApJ...759..144T} have oscillation periods of 4.13$\pm$1.46\,min and 5.35$\pm$1.29\,min, and velocity amplitudes of 1.43\,\kms and 2.73\,\kms, which are both in the range of our statistical results.

From the average velocity amplitude $\langle V\rangle $ ($1.57\, \text{km s}^{-1}$), we can calculate the average energy density by \citep{2014ApJ...795...18V}
\begin{equation}\label{key}
	\varepsilon=\frac{1}{2}(\rho_\text{i}+\rho_\text{e})\langle V\rangle^2\,.
\end{equation}
The external density $ \rho_\text{e} $ and internal density $ \rho_\text{i} $ are assumed to be $ 2\times 10^{-12} \,\text{kg m}^{-3} $ and  $ 4\times 10^{-12} \,\text{kg m}^{-3}  $ (the density ratio $ \rho_\text{i}/\rho_\text{e} $ is taken as 2 according to the results in Section \ref{subsec:seis} and Table \ref{speed}). Then we can obtain an average energy density of $ 7.39\times 10^{-6} \,\text{J m}^{-3} $.
Assuming a number density of $ 10^{15}\, \text{m}^{-3} $ and a temperature of $ 10^6 $ K, we can roughly estimate the total radiative energy loss rate $ Q $ for CBPs \cite[according to][]{2009A&A...498..915D,2020SSRv..216..136L}. The magnitude of Q is found to be $ 10^{-4} \,\text{W m}^{-3} $. Comparing it with our average energy density, we can obtain a time scale of $ 7.39\times10^{-2} $ s, which means that the energy content is much lower than what is needed.

\section{Discussion}
\label{sec:dis}

\subsection{Correlation between the oscillation parameters}
\label{subsec:cor}

\begin{figure*}[ht!]
	\centering
	\includegraphics[trim=0.0cm 0.8cm 0.0cm 0.5cm,width=1\textwidth]{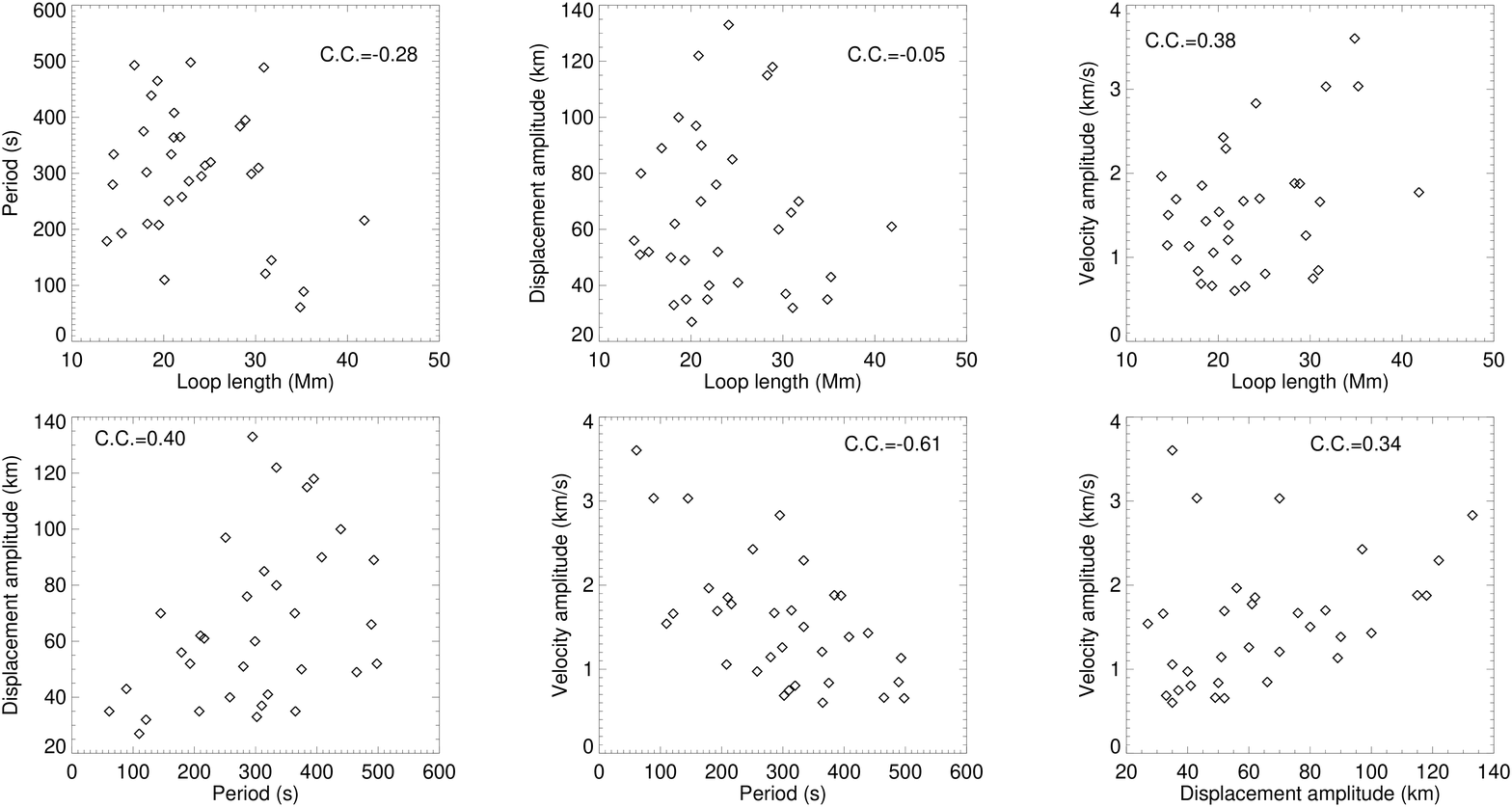}
	\caption{Scatter plots between each two of the four oscillation parameters, namely oscillation periods, displacement amplitudes, loop lengths, and velocity amplitudes. The correlation coefficients (C.C.) are indicated in each panel.}
	\label{fig:correlation}
\end{figure*}

We now investigate the relationship between the four oscillation parameters. Figure\,\ref{fig:correlation} shows the scatter plots and the correlation coefficients between each two of them.
\cite{2015A&A...583A.136A} found that the oscillation periods linearly increase with the loop lengths, which can be fitted as $ P\,(\text{s})=(1.08\pm0.04)L\,(\text{Mm}) $. 
However, as seen in Figure\,\ref{fig:correlation}, the periods and loop lengths that we obtained here have a correlation coefficient of -0.28, indicating no such correlation. 
We suggest that this could be a result of the much lower heights of CBPs than coronal loops.
The latter usually have heights of several hundred megameters. At that altitude, the kink speeds (close to the Alfv\'{e}n speeds) of different coronal loops are close to each other. For kink oscillations, the phase speed is equal to the kink speed $ C_\text{k} $ in the long-wavelength limit \citep{1983SoPh...88..179E}. So we have $ \ck\propto L/P $, which means that $ P $ is approximately proportional to $ L $ for kink oscillations in coronal loops \citep{2015A&A...583A.136A}. As for the CBPs, the loops have much lower heights. In the lower corona, the Alfv\'{e}n speeds and kink speeds could vary more in different CBPs (see Figure \ref{fig:V_Ai}). In this case, the oscillation period $ P $ will have no clear linear correlation with the loop length $ L $, just like what we observed.

There are also other possible explanations for the lack of correlation. 
Perhaps it is a bad approximation to consider the loops of CBPs as semi-circles.
These low-lying loops may have a longer horizontal part in the denser chromosphere, resulting in longer periods.
For example, the loops in Figure \ref{fig:events}(e) and (g) seem to have such a characteristic.
Another possibility is that our estimation of loop lengths may deviate from the true values. 
We used the distance between two footpoints in AIA 171\,\AA\, images to estimate the lengths.
However, the loops are rooted further down. This could introduce an error in loop 
length estimation, which is much more significant for shorter loops. 
The error could also lead to the deviation from a linear correlation.
Moreover, a constant bias in loop length calculation will result in a larger relative error compared to the measured lengths of longer loops, which means that the relative error here may be much larger than 10\% in coronal loops \cite[see][]{2007A&A...471..311V}.
Finally, we cannot rule out the possibility that the oscillation is just an externally driven displacement, rather than an eigenmode determined by the loop structure. In that case, the oscillation period just reflects the period of the driver at the footpoints, and will certainly not increase with the loop length.

From other panels of Figure\,\ref{fig:correlation}, we can see that the displacement amplitudes and periods have a weak positive correlation, while the displacements and loop lengths have no correlation. As for the velocity amplitudes, they have weak correlations with loop lengths and diplacement amplitudes. Interestingly, there seems to be a negative linear correlation between the velocity amplitudes and periods. We believe that it could be a result of our estimation method. As mentioned in Section\,\ref{sec:ana}, the velocity amplitudes are calculated with $ V=2\pi A/P $. Since the difference of $ A $ for different loops is much smaller than that of $ P $, we could expect a negative correlation between $ V $ and $ P $.

\subsection{Seismological diagnostic of Alfv\'{e}n speed}
\label{subsec:seis}

Assuming that the decayless oscillations we detected are all the fundamental mode, we could use
\begin{equation}\label{key}
	\ck=2L/P\,,
\end{equation}
to calculate the kink speeds. 
The external and internal Alfv\'{e}n speeds of the loop could be expressed as \citep[e.g.,][]{2001A&A...372L..53N,2019ApJ...884L..40A} 
\begin{equation}\label{key}
	V_\text{Ae}=\ck\sqrt{\frac{1+\rho_\text{i}/\rho_\text{e}}{2}}\,,
\end{equation}
and
\begin{equation}\label{key}
	V_\text{Ai}=\ck\sqrt{\frac{1+\rho_\text{e}/\rho_\text{i}}{2}}=\frac{V_\text{Ae}}{\sqrt{\rho_\text{i}/\rho_\text{e}}}\,,
\end{equation}
where $ \rho_\text{e} $ and $ \rho_\text{i} $ represent the external and internal densities, respectively. 
Thus, if we know the density ratio $ \rho_\text{i}/\rho_\text{e} $, we can perform seismological diagnostics of the Alfv\'{e}n speeds.

The density ratio could be estimated from the background subtracted intensity as \citep[see][]{2003ApJ...598.1375A,2007A&A...473..959V,2021ApJ...922...60S}
\begin{equation}\label{key}
	\frac{\rho_\text{i}}{\rho_\text{e}}=\sqrt{\frac{I_\text{i}-I_\text{e}}{w_\text{loop}Rn_\text{e}^2}+1}\,.
\end{equation}
Here the internal intensity $ I_\text{i} $ and the external intensity $ I_\text{e} $ are obtained from the original AIA 171\,\AA\, image by calculating the mean intensity of two sub-areas inside and outside the loop, respectively.  The loop width $ w_{\text{loop}} $ and the external number density $ n_\text{e} $ are estimated as 2\,Mm and $ 10^{15}\,\text{m}^{-3} $.  The response function $ R $ is obtained from the Solar Software. 
If we choose different sub-areas to calculate $ I_\text{i} $ and $ I_\text{e} $, the results may vary a little, but we found that the estimated density ratio barely changes. With the density ratio, we can further calculate the internal and external Alfv\'{e}n speeds.

\begin{figure*}[ht!]
	\centering
	\includegraphics[trim=0.0cm 0.8cm 0.0cm 0.5cm,width=0.5\textwidth]{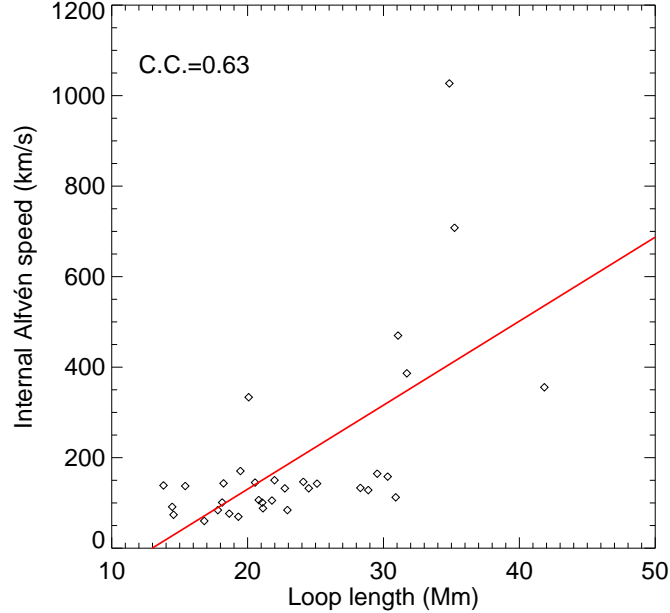}
	\caption{Scatter plots between the internal Alfv\'{e}n speeds and loop lengths. The red line represents a linear fit, and the correlation coefficient (C.C.) is also indicated.}
	\label{fig:V_Ai}
\end{figure*}

In Table\,\ref{speed}, we list all the information used for diagnosing the Alfv\'{e}n speeds, as well as the seismological results.
With a similar method, \cite{2019ApJ...884L..40A} measured the Alfv\'{e}n speeds of eight coronal loops in a quiet active region, and they found that the internal Alfv\'{e}n speeds have good accuracy. 
Considering that, here we supposed that $ V_\text{Ai} $ is also more precise.
We then made a scatter plot between $ V_\text{Ai} $ and the loop length $ L $, as shown in Figure \ref{fig:V_Ai}. 
The internal Alfv\'{e}n speed is found to increase with the loop length in a nearly linear fashion, with a correlation coefficient of 0.63.
It suggests that the Alfv\'{e}n speed is larger at a higher altitude.
In fact, the major radii of the CBP loops range from 4.5 to 13.4 Mm (calculated by $ L/\pi $). In such low heights, we would expect that the density significantly decreases with height \cite[e.g., see Figure 7 in][]{2020ApJ...894...79W}.
Therefore, a correlation between the Alfv\'{e}n speed and loop length is expected. 
Also, we note that the estimated Alfv\'{e}n speeds are much lower than those inferred from coronal loop observations (e.g., \cite{2019ApJ...884L..40A} found that the internal Alfv\'{e}n speeds of several coronal loops in a quiet solar active region are around 1000 \kms). This might be related to the low heights of CBPs and the underestimation of loop lengths.

We also made a very rough estimation of the magnetic field strength in CBPs with
\begin{equation}\label{key}
	B=V_\text{Ai}\sqrt{\mu_0 nm_\text{p}\tilde{\mu}}\,,
\end{equation}
where $ \mu_0 $ is the magnetic permeability in vaccum, $ n $ is the electron number density, $ m_\text{p} $ is the proton mass, and $ \tilde{\mu}=1.27 $ is the mean molecular weight \citep[see][]{2001A&A...372L..53N,2012A&A...537A..49W,2013A&A...552A..57N}. 
If choosing $ n=10^{15}\,\text{m}^{-3} $, we can estimate the magnetic field strength for every oscillation event, and the results are also shown in Table\,\ref{speed}. We note that the field 
strength is the average along the loop, without considering the change of density with height. For most oscillation events, the estimated magnetic fields are rather small, just around 1--3\,G. However, previous measurements of the CBP electron densities 
show a range of $ 10^{15}-10^{16}\,\text{m}^{-3} $ \citep{2019LRSP...16....2M,2021ApJ...906...59H}. 
Meanwhile, as mentioned in Section\,\ref{subsec:cor}, we are likely to underestimate the loop length, which will lead to an underestimation of the kink speed, the Alfv\'{e}n speed, and consequently the magnetic field strength.
So the real magnetic field could be approximately a few tens of Gauss. Nevertheless, to our knowledge, it is the first time that the CBP magnetic fields at the coronal heights are estimated, although the estimation is very rough with large uncertainties. 

Coronal seismology has always been seen as a prospective method of coronal magnetic field diagnostic. 
Previous studies have managed to obtain the magnetic field in active regions using standing kink or slow waves of coronal loops \cite[e.g.,][]{2001A&A...372L..53N,2007ApJ...656..598W}, and also plane-of-sky component of the magnetic field in the global off-limb corona using the propagating Alfv\'{e}nic waves \citep{2020ScChE..63.2357Y,2020Sci...369..694Y}. Our work could open up a new way to probe the
coronal magnetic field in the quiet-Sun region and coronal hole.

\subsection{How could the sub-resolution displacement amplitudes be detected?}
\label{subsec:how}

\begin{figure*}
	\centering
	\includegraphics[trim=0.0cm 0.8cm 0.0cm 0.5cm,width=1\textwidth]{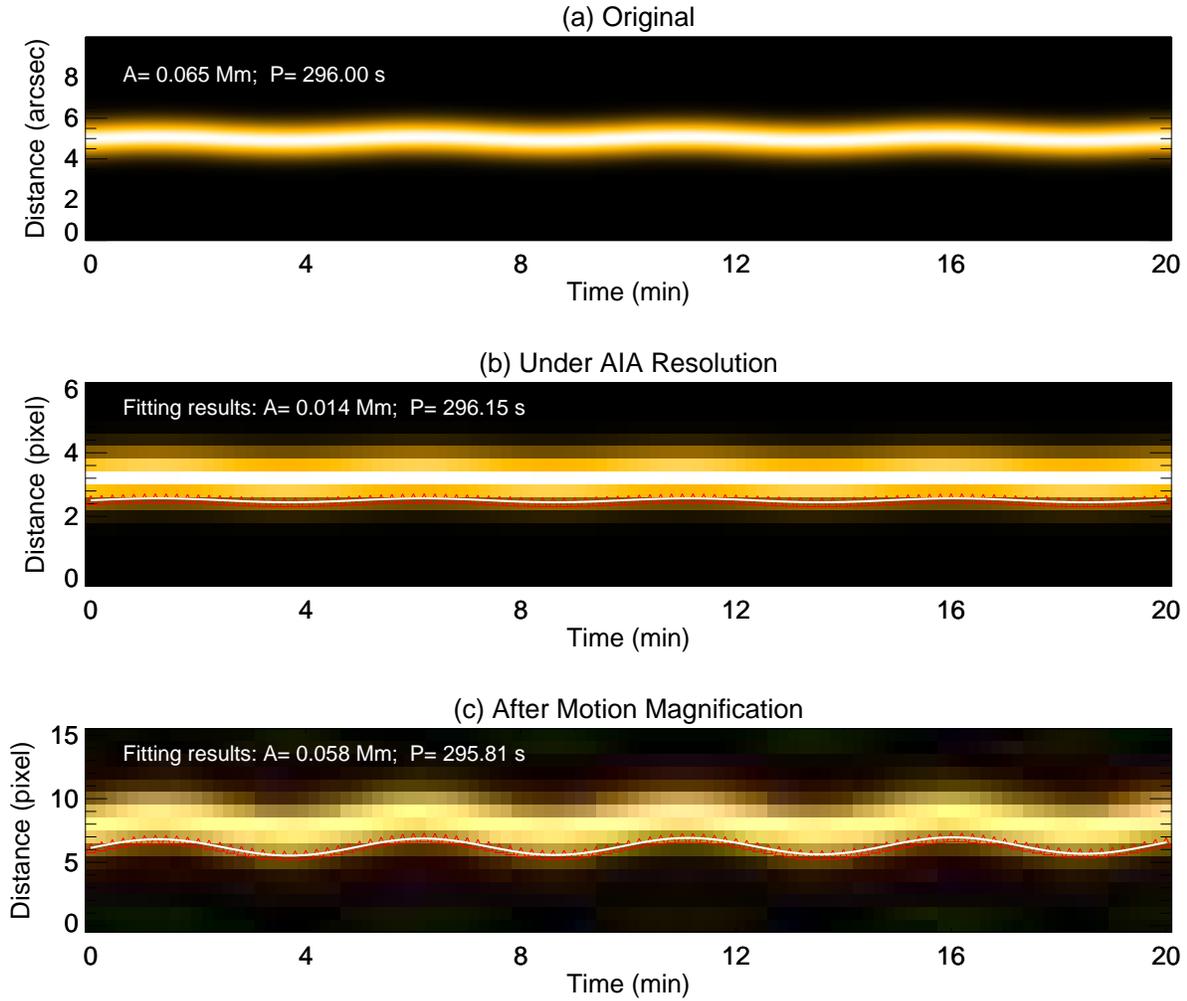}
	\caption{(a) The time-distance map made from an artificially constructed data cube with a higher spatial resolution, in which there is an oscillating loop structure with a width of 1\,Mm. The displacement amplitude $ A $ and period $ P $ of the oscillation are indicated in the upper left corner.
	(b) The time-distance map made from a data cube with a degraded resolution comparable to AIA (1.5\arcsec). The loop edges are marked with red triangles. The white curve shows the fitting result. The fitting displacement amplitude and period are also indicated in the upper left corner.
	(c) The time-distance map after motion magnification by a factor of 5, and degradation to AIA resolution. Note that the fitting amplitude indicated here has been divided by 5, like the right panels in Figure\,\ref{fig:events}.}
	\label{fig:toy1}
\end{figure*}

\begin{figure*}
	\centering
	\includegraphics[trim=0.0cm 0.8cm 0.0cm 0.5cm,width=1\textwidth]{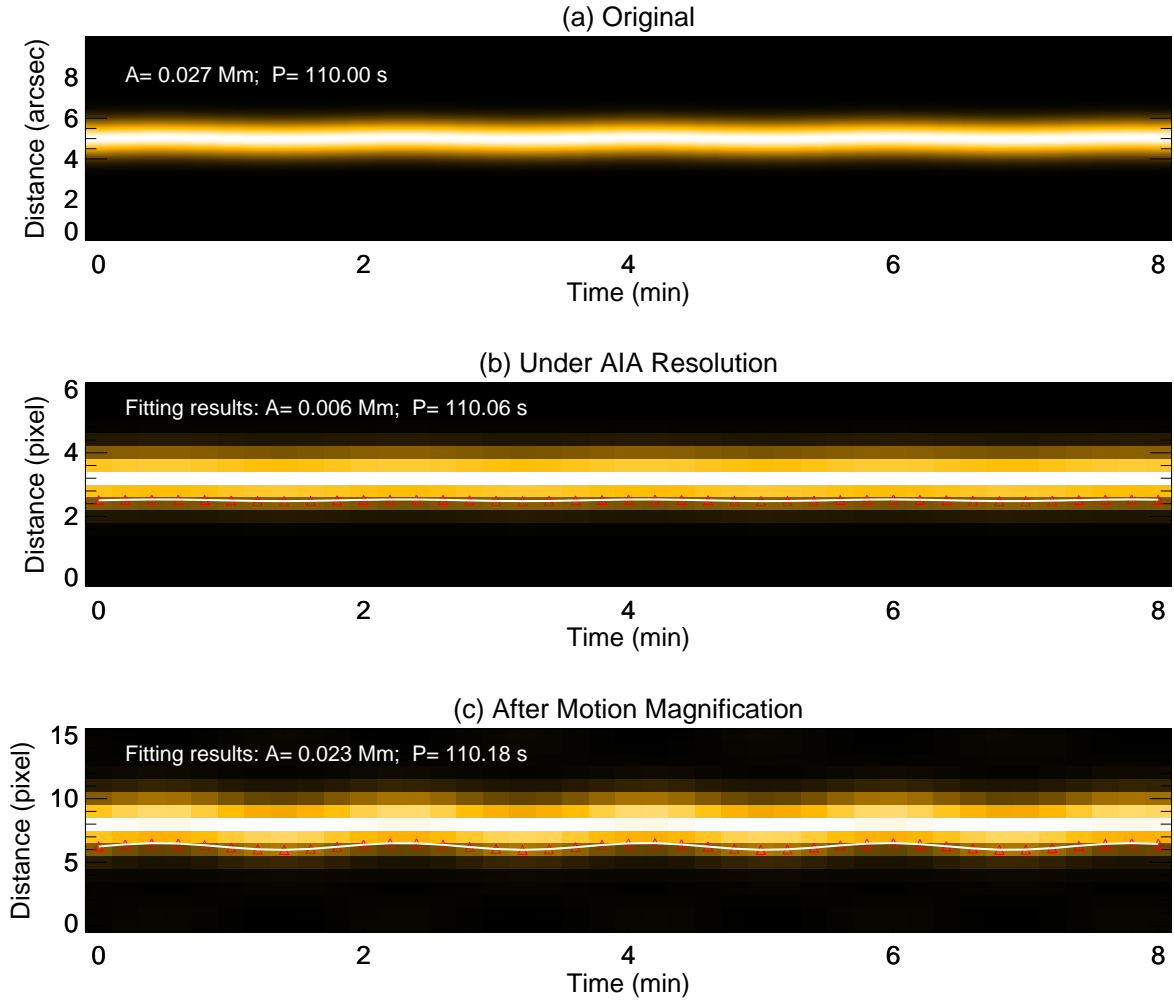}
	\caption{Similiar to Figure\,\ref{fig:toy1} but for different initial oscillation parameters ($ A=0.027\,\text{Mm};\,P=110\,\text{s} $).}
	\label{fig:toy2}
\end{figure*}

The displacement amplitudes in this study range from 27 to 133 kilometers, with an average of 65\,km, which are much smaller than the pixel size of AIA 171\,\AA\, images (around 400\,km). 
Therefore, two questions will arise: How could AIA observe these sub-pixel amplitudes? Are these parameters we obtained reliable or not?

To answer these two questions, we introduced a simple model.
We constructed an artificial data cube with three dimensions (X: solar-X, Y: solar-Y; T: time step).
There are 120 grid points in the X direction, and 1000 grid points in the Y direction.
Each grid point corresponds to 0.01\arcsec, which is 1/60 of the AIA pixel size. Therefore, the spatial range is 1.2\arcsec$\times$10\arcsec.
The time range and the cadence were chosen as 1200\,s and 12\,s, which means that our data cube has 100 frames in the time dimension.
We gave each grid point an intensity value, which is initially set to 0.
Then we added a bright loop structure with the maximum intensity value of 1 in every frame .
Because the range of the X-dimension of our data cube is only 1.2\arcsec (0.87\,Mm), we assumed that the loop is straight, and kept it parallel to the X axis. 
This means that we just modeled a very short section (0.87\,Mm) of a loop along the loop axis.
In observations, a coronal loop usually appears as a Guassian distribution in the cross-sectional intensity profile. Meanwhile, the full width at half maximum (FWHM) is taken as the loop width \citep[e.g.,][]{2012ApJ...751L..27W,2020ApJ...900..167K}. Considering that, here we also gave the artificial loop a gaussian-like intensity profile, with the FWHM of 1\,Mm. The loop center is initially set to the midpoint of the Y axis.

Next, we added a low-amplitude, decayless oscillation into the loop, letting the loop center oscillate sinusoidally with time. The oscillation parameters are taken as the observed average displacement amplitude 0.065\,Mm, and average period 296\,s.
Then, as shown in Figure\,\ref{fig:toy1}(a), a time-distance map is created along a slit perpendicular to the loop axis (or the X axis).
We can see a clear oscillation pattern at a much higher spatial resolution than AIA. 
In Figure\,\ref{fig:toy1}(b), we show the time-distance map when the spatial resolution is reduced to that of AIA (1.5\arcsec). The oscillation is still visible despite the sub-resolution amplitude. It is mainly in the form that the intensity near the loop edges changes periodically. By applying the analysis method described in Section\,\ref{sec:ana}, we can obtain the oscillation displacement amplitude and period from the fitting curve, which are 0.014\,Mm and 296.15\,s, respectively. The fitting period is close to the input value, while the fitting amplitude is much smaller.

In Figure\,\ref{fig:toy1}(c), we considered the motion magnification.
We applied the motion magnification algorithm to the original artificial data cube, and then degraded the spatial resolution.
Obviously, the time-distance map shows a magnified oscillation.
After fitting the edge of the oscillating loop, we obtained a displacement amplitude of 0.058\,Mm (about 11\% smaller than the originally setting value 0.065\,Mm), and a period of 295.81\,s.
We can see that these fitting results are in better agreement with the input values, which means that our analysis methods can give convincing results.

Additionally, we also tested other cases by choosing different initial oscillation parameters. In Figure\,\ref{fig:toy2}, we present results of the model when taking $ A $ as 0.027\,Mm, and $ P $ as 110\,s. This corresponds to the oscillation event that we observed in CBP No.\,19, which has the smallest displacement amplitude. In Figure\,\ref{fig:toy2}(b), we can see that for such a small amplitude (less than 1/10 of the AIA image pixel size), the oscillation pattern is not as obvious as in Figure\,\ref{fig:toy1}(b).
Futhermore, as we can see in Figure\,\ref{fig:toy2}(c), the fitting after motion magnification reveals an amplitude of 0.023\,Mm, which is also slightly smaller than the input amplitude, and the fitting period is close to the input value. Thus, we conclude that events with such small displacement amplitudes can also be detected, while the actual amplitudes could be slightly larger than the observed. 

After testing the oscillation parameters detected from all the oscillation events, we found that all the fitting displacement amplitudes are smaller than the input value, but the deviations are less than 20\%. Meanwhile, the fitting can provide an accurate estimation of the oscillation period.

Very recently, \cite{2021SoPh..296..135Z} investigated the motion magnification algorithm's capability and found that the algorithm works well even when analysing $ \le $ 0.01 pixel oscillations.  The algorithm also performs robustly when considering the noise. These results further enhance the reliability of our analysis methods and results.

Overall, decayless oscillations with displacement amplitudes much smaller than the AIA image pixel size could be well detected after motion magnification in the time-distance maps, and the oscillation parameters we obtained are reliable for most events.

\section{Summary}
\label{sec:sum}

In this paper, we focus on decayless oscillations in CBPs observed by the SDO/AIA 171\,\AA\, channel.
We have identified 31 oscillation events in 23 CBPs, and obtained their parameters.
Our statistical study shows that the oscillation periods are 1 to 8 minutes, with an average of about 5\,min. 
The displacement amplitudes range from 0.03 to 0.13\,Mm.
We developed a simple model to explain these sub-pixel displacement amplitudes and verify the reliability of our analysis methods.
Different from the decayless oscillations of coronal loops, no linear correlation is found between oscillation periods and loop lengths. 
In addition, coronal seismology based on
the oscillations we detected gives the kink speeds and Alfv\'{e}n speeds of CBPs, which are found to increase with height. With the Alfv\'{e}n speeds and an assumed density, we roughly estimated the CBP magnetic field strengths.

The Extreme Ultraviolet Imager \citep[EUI;][]{2020A&A...642A...8R}
onboard the Solar Orbiter \citep{2020A&A...642A...1M}
can take unprecedentedly high spatial and temporal resolution EUV images of the corona at 174\,\AA. Our analysis techniques can be applied to the EUI 174\,\AA\, data. These high-resolution observations will allow us to further investigate these decayless oscillations in CBPs in the future.

\acknowledgments
This work was supported by NSFC grants 11825301 and 11790304. AIA is an instrument on board the Solar Dynamics Observatory, a mission for NASA’s Living With a
Star program.
T.V.D. was supported by the European 
Research Council (ERC) under the European Union's Horizon 2020 research 
and innovation programme (grant agreement No. 724326) and the C1 grant 
TRACE space of Internal Funds KU Leuven.

\bibliographystyle{aasjournal}
\bibliography{bibliography}

\newpage
\begin{center}
	\renewcommand\tabcolsep{6.pt}
	\begin{longtable*}{c c c c c c c c c c c c}
		\caption{Detailed information of CBPs and decayless oscillation events}
		\label{CBPs} 
		\\
		\hline
		\multirow{2}{*}{No.} & \multirow{2}{*}{Date} & \multirow{2}{*}{Location} & Osci?& Start& End& \multirow{2}{*}{N}& $ P $& $ A $& $ L $& $ V $	\\ 
		& & &yes/no & Time & Time & & [s]& [km]& [Mm]& [\kms]]\\ \hline
		1 &
		2017-01-04 &
		E354 N323 &
		no &
		… &
		… &
		… &
		… &
		… &
		… &
		… \\ \hline
		2 &
		2017-01-17 &
		W175 S280 &
		no &
		… &
		… &
		… &
		… &
		… &
		… &
		… \\ \hline
		\multirow{4}{*}{3} &
		\multirow{4}{*}{2017-02-17} &
		\multirow{4}{*}{W99 N112} &
		\multirow{4}{*}{yes} &
		11:11:57 & 11:31:45 & 5.5 & 208$\pm$7 & 35$\pm$12 & 19.5 & 1.06$\pm$0.38\\
		&&&& 12:27:57 & 12:47:45 & 4 & 299$\pm$9 & 60$\pm$13 & 29.5 & 1.26$\pm$0.26\\
		&&&& 12:51:57 & 13:11:45 & 4 & 310$\pm$15 & 37$\pm$13 & 30.3 & 0.75$\pm$0.26\\
		&&&& 13:43:57 & 14:01:45 & 4 & 258$\pm$10 & 40$\pm$13 & 22.0 & 0.97$\pm$0.31\\
		\hline
		\multirow{2}{*}{4} &
		\multirow{2}{*}{2017-02-17} &
		\multirow{2}{*}{W249 S208} &
		\multirow{2}{*}{yes} &
		15:39:09 & 16:26:57 & 7.5 & 375$\pm$6 & 50$\pm$8 & 17.8 & 0.84$\pm$0.14 \\
		&&&&16:45:09 & 17:08:51 & 5 & 286$\pm$6 & 76$\pm$11 & 22.4 & 1.67$\pm$0.25 \\
		\hline
		5 &
		2017-03-17 &
		E304 N256 &
		no &
		… &
		… &
		… &
		… &
		… &
		… &
		… \\ \hline
		6 &
		2017-03-29 &
		E51 S108 &
		no &
		… &
		… &
		… &
		… &
		… &
		… &
		… \\ \hline
		7 &
		2017-04-28 &
		W692 N170 &
		yes &
		04:08:09 &
		04:37:57 &
		4.5 &
		408$\pm$7 &
		90$\pm$10 &
		21.1 &
		1.39$\pm$0.16 \\ \hline
		8 &
		2017-05-01 &
		W301 S301 &
		yes &
		23:50:09 &
		00:50:09 &
		7 &
		497$\pm$15 &
		49$\pm$10 &
		22.9 &
		0.62$\pm$0.13 \\ \hline
		9 &
		2017-05-24 &
		E125 S498 &
		no &
		… &
		… &
		… &
		… &
		… &
		… &
		… \\ \hline
		\multirow{2}{*}{10} &
		\multirow{2}{*}{2017-05-27} &
		\multirow{2}{*}{W170 S405} &
		\multirow{2}{*}{yes} &
		20:20:09 & 20:43:57 & 5 & 280$\pm$8 & 51$\pm$11 & 14.5 & 1.14$\pm$0.25 \\
		&&&& 21:54:09 & 22:11:57 & 6 & 179$\pm$3 & 56$\pm$13 & 13.8 & 1.97$\pm$0.46 \\
		\hline
		\multirow{2}{*}{11} &
		\multirow{2}{*}{2017-06-09} &
		\multirow{2}{*}{W435 S154} &
		\multirow{2}{*}{yes} &
		22:35:09 & 22:58:57 & 5 & 302$\pm$12 & 33$\pm$12 & 18.1 & 0.69$\pm$0.24 \\
		&&&& 23:41:09 & 00:20:57 & 5 & 465$\pm$11 & 49$\pm$9 & 19.3 & 0.66$\pm$0.11 \\
		\hline
		12 &
		2017-07-20 &
		E230 S423 &
		yes &
		21:27:09 &
		21:52:57 &
		4.5 &
		334$\pm$3 &
		122$\pm$11 &
		20.8 &
		2.30$\pm$0.20 \\ \hline
		\multirow{2}{*}{13} &
		\multirow{2}{*}{2017-08-21} &
		\multirow{2}{*}{W68 N346} &
		\multirow{2}{*}{yes} &
		21:04:57 & 21:35:45 & 3.5 & 493$\pm$8 & 89$\pm$10 & 16.8 & 1.13$\pm$0.13 \\
		&&&&23:27:57 & 23:59:45 & 4 & 439$\pm$7 & 100$\pm$10 & 18.7 & 1.43$\pm$0.15 \\
		\hline
		14 &
		2018-01-08 &
		W133 S481 &
		yes &
		03:14:09 &
		03:29:57 &
		8 &
		121$\pm$4 &
		32$\pm$14 &
		31.1 &
		1.66$\pm$0.72 \\ \hline
		15 &
		2018-02-02 &
		E119 S324 &
		yes &
		12:46:09 &
		13:15:57 &
		5 &
		365$\pm$12 &
		35$\pm$10 &
		21.8 &
		0.60$\pm$18 \\ \hline
		16 &
		2018-02-06 &
		W225 S393 &
		yes &
		11:36:09 &
		12:05:57 &
		5.5 &
		314$\pm$4 &
		85$\pm$10 &
		24.5 &
		1.70$\pm$0.20 \\ \hline
		\multirow{4}{*}{17} &
		\multirow{4}{*}{2018-02-17} &
		\multirow{4}{*}{W490 N55} &
		\multirow{4}{*}{yes} &
		16:28:09 & 16:55:57 & 4.5 & 364$\pm$7 & 70$\pm$10 & 21.1 & 1.21$\pm$0.18 \\
		&&&& 18:04:09 & 18:23:57 & 5 & 210$\pm$4 & 62$\pm$13 & 18.2 & 1.86$\pm$0.38 \\
		&&&& 19:40:09 & 19:56:57 & 3 & 334$\pm$4 & 80$\pm$10 & 14.6 & 1.51$\pm$0.19 \\
		&&&& 19:56:57 & 20:09:57 & 3.5 & 193$\pm$2 & 52$\pm$10 & 15.4 & 1.69$\pm$0.33 \\
		\hline
		\multirow{2}{*}{18} &
		\multirow{2}{*}{2018-03-04} &
		\multirow{2}{*}{E253 N274} &
		\multirow{2}{*}{yes} &
		17:08:09 & 17:13:57 & 5.5 & 61$\pm$4 & 35$\pm$23 & 34.9 & 3.61$\pm$2.36 \\
		&&&& 18:16:09 & 18:33:57 & 7 & 145$\pm$2 & 70$\pm$13 & 31.7 & 3.03$\pm$0.57 \\
		\hline
		\multirow{5}{*}{19} &
		\multirow{5}{*}{2018-03-07} &
		\multirow{5}{*}{W685 N254} &
		\multirow{5}{*}{yes} &
		01:16:09 & 01:23:57 & 4 & 110$\pm$10 & 27$\pm$20 & 20.1 & 1.54$\pm$1.15 \\
		&&&& 01:26:09 & 01:41:57 & 3.5 & 251$\pm$5 & 97$\pm$14 & 20.6 & 2.43$\pm$0.35 \\
		&&&& 02:50:09 & 03:10:57 & 4 & 295$\pm$3 & 133$\pm$12 & 24.1 & 2.83$\pm$0.27 \\
		&&&& 03:20:09 & 03:59:57 & 7 & 320$\pm$5 & 41$\pm$9 & 25.1 & 0.81$\pm$0.17 \\
		&&&& 05:06:09 & 05:29:57 & 4 & 384$\pm$6 & 115$\pm$12 & 28.3 & 1.88$\pm$0.19 \\
		\hline
		20 &
		2018-03-11 &
		W398 N97 &
		no &
		… &
		… &
		… &
		… &
		… &
		… &
		… \\ \hline
		21 &
		2018-03-14 &
		W34 S21 &
		no &
		… &
		… &
		… &
		… &
		… &
		… &
		… \\ \hline
		22 &
		2018-03-16 &
		E31 S94 &
		yes &
		00:40:09 &
		01:05:57 &
		4 &
		395$\pm$5 &
		117$\pm$11 &
		28.9 &
		1.88$\pm$0.18 \\ \hline
		23 &
		2018-03-25 &
		E181 N389 &
		yes &
		15:18:09 &
		15:35:57 &
		5 &
		216$\pm$5 &
		61$\pm$13 &
		41.8 &
		1.77$\pm$0.39 \\
		\hline
	\end{longtable*}
	\tablenotetext{}{\textbf{Notes.} The following information is listed for each event: 
		ID of the CBPs, 
		observation date (yyyy-mm-dd),
		coordinates of the location, 
		whether decayless oscillation events exist or not (yes/no), 
		starting time (hh:mm:ss), 
		end time (hh:mm:ss),
		cycle number (N),
		period ($ P $, s), 
		displacement amplitude ($ A $, km),
		loop length ($ L $, Mm),
		velocity amplitude ($ V $, \kms).
		The  uncertainties of the periods, displacement amplitudes and velocity amplitudes are also indicated.
		}
\end{center}

\begin{center}
	\renewcommand\tabcolsep{6.pt}
	\begin{longtable*}{c c c c c c c c c c}
		\caption{Results of the coronal seismology}
		\label{speed} \\
		\hline
		\multirow{2}{*}{Date} & Start& End& $ \ck $& $ I_\text{i} $& $ I_\text{e} $& 	\multirow{2}{*}{$ \rho_\text{i}/\rho_\text{e} $}& $ V_\text{Ai} $&$ V_\text{Ae} $&$ B $	\\ 
		& Time & Time&[\kms]&[DN]&[DN]& & [\kms]&[\kms]&[G]\\ \hline
		2017-02-17               & 11:11:57                 & 11:31:45                         & 187.2                        & 514                  & 181                  & 1.52              & 170.6                    & 209.9                    & 2.78                \\ \hline
		2017-02-17               & 12:27:57                 & 12:47:45                         & 197.6                          & 1954                  & 505                  &2.58              & 164.6                    & 264.2                      & 2.69               \\ \hline
		2017-02-17               & 12:51:57                 & 13:11:45                         & 195.5                         & 2865                 & 421           & 3.24              & 158.2                    & 284.7                   & 2.58              \\ \hline
		2017-02-17               & 13:43:57                 & 14:01:45                         & 170.4                          & 855                  & 278                  & 1.80              & 150.2                     & 201.7                    & 2.45                \\ \hline
		2017-02-17               & 15:39:09                 & 16:26:57                         & 95.0                          & 914                  & 377                  & 1.76              & 84.1                    & 111.5                    & 1.37                \\ \hline
		2017-02-17               & 16:45:09                 & 17:08:57                         & 159.0                           & 1885                  & 383                  & 2.62              & 132.2                    & 213.8                   & 2.16               \\ \hline
		2017-04-28               & 04:08:09                  & 04:37:57                          & 103.6                          & 1241                  & 229                  & 2.22              & 88.2                    & 131.5                    & 1.44             \\ \hline
		2017-05-01               & 23:50:09                 & 00:50:09                          & 92.3                          & 492                  & 183                  & 1.48               & 84.2                   & 102.6                    & 1.38                \\ \hline
		2017-05-27               & 20:20:09                 & 20:43:57                         & 103.2                          & 817                 & 271                  & 1.77            & 91.3                   & 121.4                    & 1.49               \\ \hline
		2017-05-27               & 21:54:09                 & 22:11:57                         & 154.3                        & 642               & 219                  & 1.63              & 138.6            & 176.8                 & 2.26                 \\ \hline
		2017-06-09               & 22:35:09                 & 22:58:57                         & 120.1                          & 1532                  & 297                  & 2.41              & 101.0                    & 156.7                    & 1.65                \\ \hline
		2017-06-09               & 23:41:09                 & 00:20:57 							& 83.1                          & 1558                  & 242                   & 2.47             & 69.6                    & 109.5                   & 1.14                \\ \hline
		2017-07-20               & 21:27:09                 & 21:52:57                         & 124.7                          & 1220                  & 246                  & 2.19              & 106.4                    & 157.4                    & 1.74                \\ \hline
		2017-08-21               & 21:04:57                 & 21:35:45     						& 68.2                          & 732                  & 159                  & 1.80              & 60.2                    & 80.6                    & 0.98                \\ \hline
		2017-08-21               & 23:27:57                 & 23:59:45     						& 85.0                          & 579                  & 164                  & 1.62               & 76.4                    & 97.2                    & 1.25                 \\ \hline
		2018-01-08               & 03:14:09                  & 03:29:57                          & 513.6                           & 479                 & 171                   & 1.48              & 469.9                    & 572.2                    & 7.67                \\ \hline
		2018-02-02               & 12:46:09                 & 13:15:57                         & 119.4                          & 759                   & 192                  & 1.79              & 105.4                    & 141.0                    & 1.72               \\ \hline
		2018-02-06               & 11:36:09                 & 12:05:57                         & 156.1                          & 1489                  & 411                  & 2.28               & 132.4                    & 199.8                    & 2.16                \\ \hline
		2018-02-17               & 16:28:09                 & 16:55:57                         & 115.8                          & 891                  & 177                 & 1.94              & 100.8                     & 140.5                    & 1.65                \\ \hline
		2018-02-17               & 18:04:09                 & 18:23:57                         & 173.6                          & 1922                  & 214                  & 2.76              & 143.3                    & 238.2                    & 2.34                \\ \hline
		2018-02-17               & 19:40:09                 & 19:56:57                         & 87.15                           & 1294                     & 211                  & 2.28             & 73.9                    & 111.6                    & 1.21               \\ \hline
		2018-02-17               & 19:56:57                 & 20:09:57                         & 159.7                          & 1068                  & 210                  & 2.08            & 137.4                    & 198.3                   & 2.24                \\ \hline
		2018-03-04               & 17:08:09                 & 17:13:57                         & 1142.61                          & 640                  & 219                  & 1.62              & 1027.0                    & 1308.9                    & 16.8                 \\ \hline
		2018-03-04               & 18:16:09                 & 18:33:57                         & 437.5                          & 795                   & 233                  & 1.78              & 386.4                    & 516.3                    & 6.31                \\ \hline
		2018-03-07               & 01:16:09                  & 01:23:57                          & 365.1                          & 467                 & 152                  & 1.49              & 333.6                    & 407.5                    & 5.45               \\ \hline
		2018-03-07               & 01:26:09                  & 01:41:57                          & 163.7                          & 696                  & 161                  & 1.75              & 145.1                    & 192.2                    & 2.37                \\ \hline
		2018-03-07               & 02:50:09                  & 03:10:57                          & 163.4                          & 654                  & 220                  & 1.64              & 146.6                    & 187.7                     & 2.39                \\ \hline
		2018-03-07               & 03:20:09                  & 03:59:57                          & 156.9                             & 561                  & 218                  & 1.53             & 142.7                    & 176.4                   & 2.33                \\ \hline
		2018-03-07               & 05:06:09                  & 05:29:57                          & 147.4                          & 578                  & 193                  & 1.58              & 133.2                    & 167.4                   & 2.17                \\ \hline
		2018-03-16               & 00:40:09                  & 01:05:57                          & 146.2                         & 964                  & 345                  & 1.85              & 128.4                   & 174.4                    & 2.10                \\ \hline
		2018-03-25               & 15:18:09                 & 15:35:57                         & 387.4    							& 468                  & 177                  & 1.46             & 355.5                    & 429.7                   & 5.80         \\  
		\hline
	\end{longtable*}
	\tablenotetext{}{\textbf{Notes. }The following information is listed for each oscillation event: 
		observation date (yyyy-mm-dd),
		starting time (hh:mm:ss), 
		end time (hh:mm:ss),
		kink speed ($ \ck $, \kms),
		internal intensity ($ I_\text{i} $, DN), 
		external intensity ($ I_\text{e} $, DN),
		density ratio ($ \rho_\text{i}/\rho_\text{e} $),
		internal Alfv\'{e}n speed ($ V_\text{Ai} $, \kms), 
		external Alfv\'{e}n speed ($ V_\text{Ae} $, \kms),
		magnetic field ($ B $, G).}
\end{center}

\end{document}